\documentclass[11pt,onecolumn]{article}
\usepackage{epsfig,graphics,graphicx,latexsym,amsmath,caption}
\begin{document}

\pagestyle{plain}

\begin{center}

\vspace*{-1.2cm}

\vspace*{2.0cm}

\vspace {0.6cm}

{\huge\bf Baryon number in proton-proton \\ 
\vspace*{0.2cm}
and proton-nucleus high energy collisions}

\vspace {0.7cm}
\vspace {0.2cm}

{\large {Marek Je\.zabek$^1$ and Andrzej Rybicki$^1$}}

\vspace {0.2cm}
\vspace {0.4cm}
{$^1$Institute of Nuclear Physics, Polish 
 Academy of Sciences, \\
Radzikowskiego 152, 31-342~Krak\'ow, Poland}
\vspace {0.2cm}
\vspace {0.4cm}

{marek.jezabek@ifj.edu.pl\\
\vspace {0.0cm}
andrzej.rybicki@ifj.edu.pl\\}

\vspace {0.2cm}
\vspace {0.4cm}

{\em version 1.0, 28 May 2021}\\


\vspace {2.2cm}

{\bf Abstract}

\end{center}

\noindent
New analyses of baryon spectra in proton-proton and proton-carbon collisions at $\sqrt{s}_\mathrm{_{NN}}=17.3$ GeV, made in the framework of two phenomenological models are presented. The first model in question is the classic Dual Parton Model by Capella and Tran Thanh Van, the second is the Gluon Exchange Model very recently proposed by the authors. For both studies, the usage of modern experimental data from the CERN SPS eliminates several of the most important limitations inherent to earlier studies of this type. In both studies, the standard mechanism of baryon stopping with preservation of the diquark, proposed by Capella and Tran Thanh Van fails to describe the distribution of non-strange baryons in collisions of the projectile proton with {\em more than one} nucleon from the carbon target obtained from experimental data, and the upper limit for the contribution of this mechanism can be established. In both cases, the conclusion is that the projectile diquark must be very often disintegrated. This opens new diagrams not available in proton-proton collisions which lead to the transport of baryon number over long distances in rapidity. The present limitations, and possibility of improvement in both approaches are discussed. The implications of our findings for new measurements are addressed, in particular using antiproton beams.

\newpage

\section{Introduction}
\label{sec1}

The process of transport of baryon number from the initial to the final state plays a special role in studies of non-perturbative (``soft'') processes, induced by the strong interaction. Baryon number conservation implies that at least at not too high collision energies where baryon-antibaryon pair production is reasonably limited, baryon spectra can provide a cleaner way to investigate the fate of quarks in the reaction than, {\em e.g.}, spectra of produced particles. Therefore, they give a better tool to constrain the inherent dynamical scenarios. Also, in heavy ion collisions, baryon number transport plays a fundamental role as it lies at the basis of the energy deposit necessary to create a deconfined quark-gluon plasma~\cite{buszaledoux}.

	Seen in this context, the state of the art knowledge on the dynamics of this process appears as quite incomplete, and not exempt from spectacular controversies~\cite{bialasbzdak}. This situation we attribute at least partially to the limitations inherent to 
experimental data which originally served to build the core of the present understanding of baryon stopping 
phenomena, see {\em e.g.}~\cite{Brenner,CapellaTranh,MJ1985,Busza}. 
Because of the above-mentioned problem of pair production, and for experimental reasons connected to the lack of coverage in the longitudinal direction up to high values of $x_F$, no much hope for a significant improvement in this specific domain can be expected from LHC experiments alone; this is notwithstanding very valuable measurements emerging in the framework of the AFTER@LHC program~\cite{a}. Consequently, the question arises whether new insight into the baryon stopping process could be gained at collision energies lower than the LHC, but still belonging to the high energy, ultrarelativistic domain. 

	In this paper we present a synthetic discussion of two new analyses of transport of baryon number in proton-proton and proton-nucleus collisions. Both studies were based on modern experimental data on proton and neutron spectra in $pp$ and $pC$ collisions obtained at the CERN SPS at beam energy of 158 GeV ($\sqrt{s}_\mathrm{_{NN}}$=17.3 GeV)~\cite{x,y}. These analyses were made in the framework of two phenomenological models (1) the Dual Parton Model (DPM) proposed by Capella and Tran Thanh Van~\cite{CapellaTranh}, which makes it partially similar to earlier studies made by one of us~\cite{MJ1985,MJZalewski}, and (2) the Gluon Exchange Model (GEM) which we proposed very recently and which can be considered, technically, as an extension of the DPM with a natural, even if very significant, broadening of the available Fock space. Some of the principal results from both analyses were separately published in Refs.~\cite{e} and~\cite{g}. The aim of the present paper is to discuss important results not included therein as well as point out specific limitations of both approaches. The discussion of possible improvements of our approach (2) will bring up the importance of isospin effects in baryon transport studies, up to now poorly known due to the poor availability of neutron data.

Here, we underline that the usage of the very high quality,
cited experimental data lifts up the most important limitations inherent to earlier studies of this type. In particular, the coverage of the full proton ($x_F$,$p_T$) hemisphere allows for a precise evaluation of ``diffractive'' and ``non-diffractive'' processes, and no doubtful assumptions on the ``isospin flip'' into neutrons are needed anymore as the latter can be estimated from experimental data. Additionally, the fact that both $pp$ and $pC$ data sets were provided by the same experiment allows for the isolation of baryon spectra in {\em multiple} proton-nucleon collisions, as it will be described in Sec.~\ref{mult}. As a result, stronger and more reliable conclusions can be formulated on different aspects of the phenomenon of transport of baryon number from the initial to the final state.

The remainder of this paper is organized as follows. In Sec.~\ref{exp} we discuss the experimental data~\cite{x,y}, with special attention on advantages which these data bring to baryon transport studies. In Sec.~\ref{dpm} we review our analysis performed in the framework of the Dual Parton Model. Our new Gluon Exchange Model is presented in Sec.~\ref{gemana}, with emphasis on new results not reported earlier in Ref.~\cite{g}. A synthetic discussion, focussed on the main conclusions emerging from both studies, is made in Sec.~\ref{sum}.

\section{The Experimental Progress}
\label{exp}

The data on $pp$ and minimum bias $pC$ collisions were obtained by the NA49 experiment~\cite{nim}. For both reactions the proton data cover the entire projectile hemisphere (up to $x_F=0.95$ and with no lower cutoff in transverse momentum), while the neutron data start at $x_F=0.1$. Antiproton measurements are available in the entire region of the latter hemisphere where antiproton production is significant, at least up to $x_F=0.3$. Proton and antiproton data are available as a dense double differential grid as a function of ($x_F$, $p_T$), supplemented by a two dimensional numerical interpolation of the data points provided by the authors, as well as a rapidity distribution. For neutrons only the $x_F$ distribution is available.

	The integration of NA49 proton and neutron spectra over the entire projectile hemisphere of $pp$ collisions, supplemented by a compilation of strange baryon multiplicities provided in Ref.~\cite{Varga2003}, gives us unity within an accuracy better than 5\%. This is fully compatible with statistical and systematic errors published in Ref.~\cite{x}.

	In order to obtain the neutron rapidity distribution necessary for our studies, we assume the shape of the neutron $p_T$ spectrum at a given $x_F$ is similar to that of the proton $p_T$ distribution. The resulting ``net'' neutron $dn/dy$ rapidity distribution computed from NA49 $pp$ data is presented in Fig. 1 in comparison to the published ``net'' proton spectrum. We proceed in a similar way for $pC$ collisions.
We underline that in order to 
eliminate the contribution from baryon-antibaryon pair production, we always deal with net proton ($p-\overline{p}$) and net neutron ($n-\overline{n}$) spectra. For simplicity, we apply the assumption $\overline{n}\approx\overline{p}$
when obtaining the latter from experimental data.



\begin{figure}[t]
\centering 
\hspace*{-1cm}
\includegraphics[width=8cm]{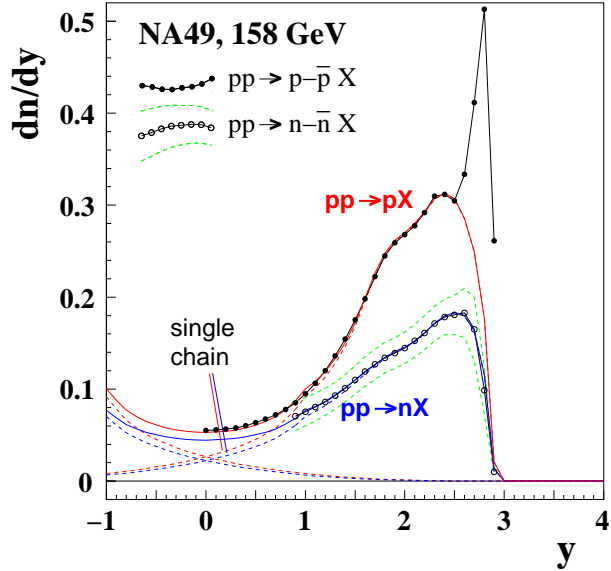}\\
\caption{{ Rapidity distributions of net protons and net neutrons in $pp$ collisions at $\sqrt{s}_\mathrm{_{NN}}=17.3$~GeV obtained from the NA49 experiment~\cite{x} (data points), put together with the result of our model simulation (solid lines). The contributions of the single $D-q$ and $q-D$ chains to the total neutron and proton spectra are marked as dashed lines. The two green dashed curves reflect the systematic error of the NA49 neutron data which is the main source of uncertainty in the paper~\cite{x}.}}
 \label{fig1}
\end{figure}

\section{The Dual Parton Model Analysis}
\label{dpm}

\begin{figure}
\centering
\begin{center}
\includegraphics[width=11cm]{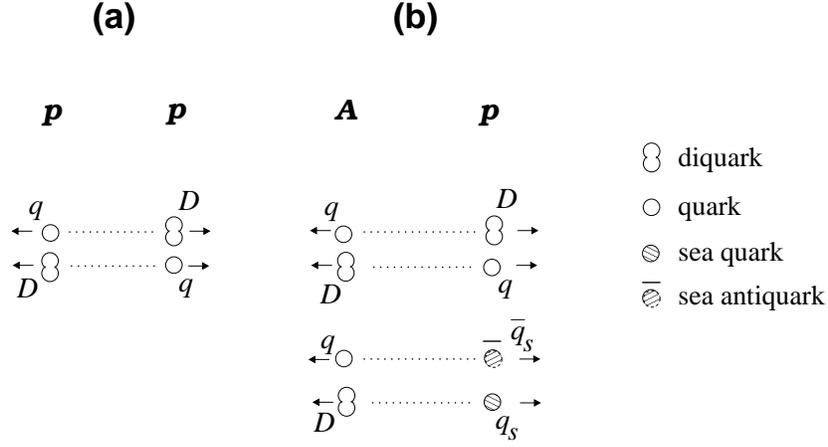}\\
\end{center}
  \caption{DPM diagrams for (a) $pp$ and (b) $pA$ collisions with $n=2$. 
A similar 
way of plotting these diagrams was 
earlier
introduced by us in Ref.~\cite{g}.
 \label{fig1314}}
\end{figure}

An exhaustive description of the Dual Parton Model (DPM) can be found in the original work~\cite{CapellaTranh}. A detailed account of its application to baryon spectra is given in Ref.~\cite{MJ1985}, and an additional discussion in the context of the color quantum number can be found 
in our recent work~\cite{e}. Only a very concise description will be made here. The DPM assumes the inelastic non-perturbative $pp$ collision to occur via color (gluon) exchange between the projectile and target protons, which results in the formation of two new color singlets, as illustrated in Fig.~\ref{fig1314}~(a). Each of these singlets (labeled ``chains'' or ``strings'') contains a diquark $D$ and a quark $q$. In the present work we will follow our nomenclature introduced in Ref.~\cite{e} and label $D-q$ the singlet made of the projectile diquark and target quark, and $q-D$ that made of the projectile quark and target diquark. The function defining the subsequent fragmentation of the latter singlets into baryons must be obtained from experimental data. This was attempted in the past~\cite{MJ1985} and was again performed, with better accuracy, in the present work.

	For the proton-nucleus collision where the incoming proton interacts with multiple nucleons, the original diagram proposed by Capella and Tran Thanh Van (which we will address as ``the standard mechanism'') is shown in Fig.~\ref{fig1314}~(b)
for scattering on two nucleons in the nucleus.
Generally, the
projectile undergoing $n$ collisions is composed of $2n$ partons, out of which $2n-2$ are sea quarks and antiquarks. As a result $n+1$ chains 
with net baryon number
are formed, all of them of the diquark-quark type just as it was the case for $pp$ collisions\footnote{Obviously, $n-1$ chains of the valence quark-sea antiquark type are also formed, but these contain no net baryon number.}. The fragmentation of the $D-q$ chain containing the projectile proton diquark occurs via the same fragmentation function as in $pp$ collisions, bringing no additional free parameters to the model.

	In the first part of our work (Secs.~\ref{pp}-\ref{bstop}) we will directly apply the complete mathematical formalism formulated earlier in Ref.~\cite{MJ1985}, which we developed in Ref.~\cite{e} into a Monte Carlo code including not only $D-q$ but also $q-D$ singlets; the only exception to the above will be fragmentation functions into protons and neutrons which can be defined far more reliably with the new experimental data. The further extension of our approach in the second part of our work will be introduced in Secs.~\ref{qend}-\ref{alt}.

\begin{figure}
\centering
\hspace*{-1cm}
\includegraphics[width=6.5cm]{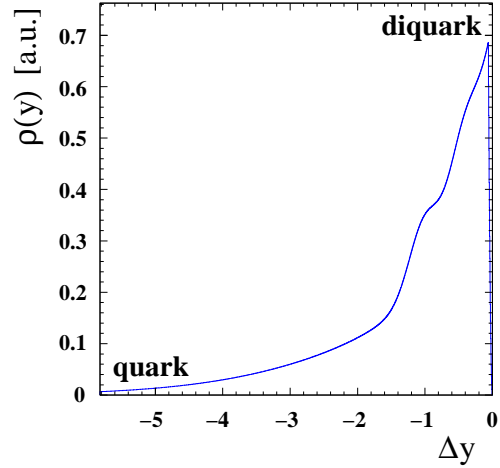}\\
  \caption{Fragmentation function of the $D-q$ (diquark-quark) chain into baryons, drawn {\em versus} the rapidity difference $\Delta y=y-y_\mathrm{max}$, where $y$ is the rapidity of the secondary baryon $B$ and $y_\mathrm{max}$ is the kinematical limit for $B$ in the direction of $D$.}
 \label{fragm}
\end{figure}

 \subsection{Baryon Spectra in $pp$ Collisions}
\label{pp}

Fig.~\ref{fig1} presents the description of net proton and net neutron spectra in $pp$ reactions, which we obtained in the framework of the DPM. The model achieves an essentially precise description of both distributions, with the evident exception of the ``diffractive'' proton peak at high rapidity. We note that the latter peak was generally believed not to be caused by color exchange processes, see {\em e.g.} Ref.~\cite{9x}, an opinion which was only challenged by the GEM model~\cite{g} which will be discussed in Sec.~\ref{gemana}. 

The shape of the fragmentation function of the $D-q$ chain into non-strange baryons is presented in Fig.~\ref{fragm}. 
Account taken of the purely effective character of this function,
enforced in detail by the experimental data, we do not bring any specific physical significance to the local structures visible in the Figure.  
We also note that this shape preserves only a basic similarity to that postulated in Ref.~\cite{MJ1985} on the basis of less precise data. Evidently the fragmentation of the $q-D$ chain will be defined by the same (reflected) fragmentation function.


\begin{figure}
\begin{center}
\hspace*{-1cm}
\includegraphics[width=6.5cm]{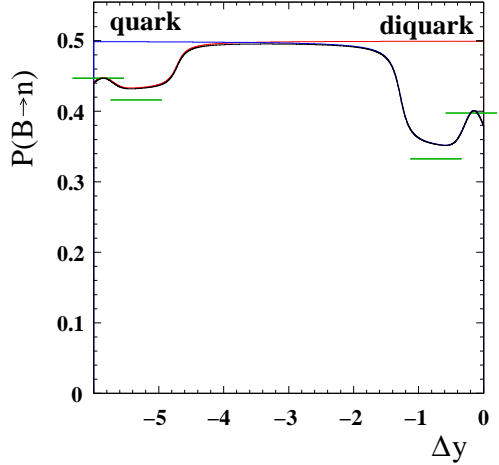}\\
\end{center}
  \caption{Isospin flip function which gives the probability $P(B\rightarrow n)$ of the non-strange baryon to become
a neutron (solid black curve). The small green 
lines illustrate our estimates obtained from 
simple quark counting. The red and
blue curves will be discussed in 
Sec.~\ref{qend}.}
 \label{isoflip}
\end{figure}

Fig.~\ref{isoflip} shows the isospin flip function, which defines the probability of a non-strange baryon to become a neutron rather than a proton. This function is also adjusted to experimental $pp$ data and is supposed to describe, in a purely effective way, the process of fragmentation of the $D-q$ chain. Nevertheless, the data enforce quite a characteristic shape for this function, which reflects the importance of isospin effects induced by the valence (uud) structure of the proton. Consequently, the diquark is composed of (uu) or (ud) quarks, while the valence quark from the opposite incoming proton is twice more often a u than a d quark. The fragmentation of the chain into a baryon in close vicinity of the rapidity of the diquark ($\Delta y\approx 0$ in the figure) is therefore characterized by a relatively low probability to form a neutron (udd), and a higher probability to form a proton (uud). When moving further away in rapidity along the string, a (natural) equilibration of both probabilities takes place, which results in the isospin flip function reaching 0.5 in this region of $\Delta y$. This fully corresponds to the proton and neutron data at high rapidity, Fig. 1, and in fact leaves no freedom for the corresponding effective shape of the isospin flip function in the latter region. 

On the other hand, what is {\em not} enforced by the experimental data is the {\em quark} end of the string. This is due to the lack of experimental neutron data~\cite{x} in the region $x_F<0.1$, see Fig.~\ref{fig1} for comparison, and additionally to the increase of the systematic uncertainty of the latter when approaching this limit. For the studies made in Refs.~\cite{e,g} as well as for the present paper, we adopt a reasonably natural assumption of a basically similar - in terms of shape as a function of rapidity - isospin effect at the diquark end {\em and} the quark end of the string. The respective importance of these two effects, apparent in Fig.~\ref{isoflip}, we estimate from basic quark-counting arithmetics.
We note that the presence of isospin effects at both ends of the string has possible implications for the role of the latter in different areas of phase space, a subject which we will address in Sec.~\ref{qend}.

 \subsection{Collision of a Proton with More than One Nucleon}
\label{mult}

The second step of our study is the analysis of proton-carbon collisions. Here
we apply the phenomenology already presented in Refs.~\cite{e,g} which can be referred to for comparison. We use a Glauber simulation which one of us published elsewhere~\cite{pcdiscus} in order to subdivide the inclusive inelastic $pC$ sample provided by NA49~\cite{y} into proton-carbon events where the projectile proton collides with a single nucleon, and these in which the proton collides with multiple nucleons. We denominate the respective probabilities of the two event classes as $P(1)$ and $1-P(1)$ respectively and note from Ref.~\cite{pcdiscus}, that the majority of  inclusive inelastic $pC$ events are in fact single proton-nucleon collisions. Following a reasoning identical to that made therein, we state that the latter will be similar to $pp$ collisions and that consequently, the net baryon ($B-\overline{B}$) distribution in proton-carbon events where the proton collides with more than one nucleon will be given by the following simple equation:
%
\begin{multline}
\frac{dn}{dy}\left(pC_\mathrm{multiple~collisions}\rightarrow(B-\bar{B})X\right)=\\
\frac{1}{1-P(1)}~
\biggl(~
\frac{dn}{dy}\left(pC\rightarrow(B-\bar{B})X\right)
-
P(1)\cdot
\frac{dn}{dy}\left(pp\rightarrow(B-\bar{B})X\right)~\biggr)~~~
\label{eq1}
\end{multline}

\begin{figure*}[t]
\begin{center}
\hspace*{-1.3cm}
\includegraphics[width=15cm]{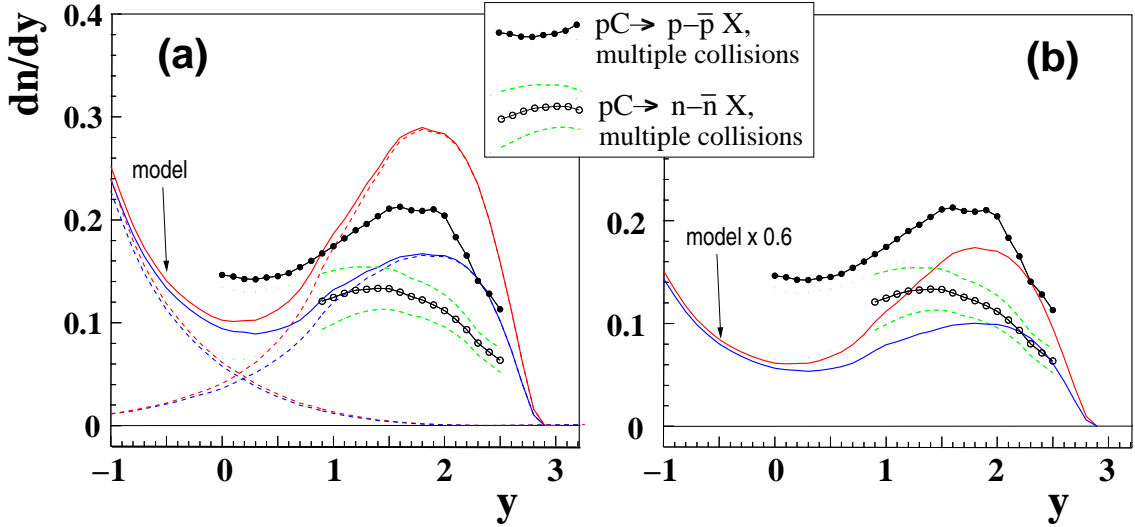}\\
\end{center}
  \caption{(a) Rapidity distributions of net protons and net neutrons in $pC$ reactions in which the projectile proton undergoes more than one collision with carbon target nucleons, obtained from the NA49 experimental data~\cite{x,y} and compared to our DPM simulation assuming the standard mechanism by Capella and Tran Thanh Van as described in the text (solid curves). The dashed curves show the 
contributions of the $D-q$ and $q-D$ chains to the total distribution. (b) The same data points compared to the result of the same simulation scaled by 0.6.  In both figures, the two green dashed curves reflect the systematic error of the NA49 neutron data.}
 \label{pcpn}
\end{figure*}

 \subsection{Baryon Stopping in the Collision of a Proton with Multiple Nucleons}
\label{bstop}

Presently we come to 
the comparison of our DPM predictions with the baryon spectra in multiple proton-nucleon collisions obtained from experimental data~\cite{x,y} with the help of Eq.~(\ref{eq1}). The study made for the total non-strange baryon spectrum, that is, the summed distribution of net protons and net neutrons, has been already presented in Ref.~\cite{e}. Here we extend this analysis to protons and neutrons taken separately and show that it brings the same conclusions as in the cited paper.

In Fig.~\ref{pcpn} we present the distributions of net protons and net neutrons in $pC$ reactions in which the projectile proton collides with more than one nucleon, 
obtained through Eq.~(\ref{eq1}). They are compared to the corresponding DPM simulation of the multiple collision process, which applies the standard mechanism by Capella and Tran Thanh Van which we presented in Fig.~\ref{fig1314} (b). We remind that the implementation of this mechanism brings no additional free parameters to the model (specifically we underline that the $D-q$ chain fragmentation and isospin flip function remain the same). This makes the model's prediction unequivocal, in particular in view of its agreement with $pp$ data apparent in Fig.~\ref{fig1}. 

	It is evident from Fig.~\ref{pcpn} that this prediction, marked as solid curve, does not match the data. The standard mechanism of softening of the projectile $D-q$ chain underpredicts the nuclear stopping power~\cite{Busza} in processes where the proton collides with more than one nucleon from the carbon nucleus. The observed difference is very large (we note that this way an old controversy~\cite{comments} is fixed in favor of Ref.~\cite{Busza}). As a result, we conclude that in the collision of the projectile proton with $n>1$ nucleons, the standard scenario of $D-q$ chain fragmentation {\em cannot be considered as the only mechanism which brings the baryon number from the initial to the final state}.

       The precision of the experimental data allows us to establish a tentative upper limit for the contribution of this standard mechanism to the net proton and net neutron spectra. This is presented in Fig.~\ref{pcpn}(b), which shows our model simulation scaled by 0.6. 
Account taken of experimental uncertainties, this would correspond to an upper limit of about two thirds
for the standard diquark-preserving mechanism.
This is consistent with what 
we obtained on the basis of total net baryon spectra in Ref.~\cite{e}. As we said therein, the remaining part must be attributed to some other mechanism, the importance of which is expected to increase for baryons found at lower rapidities.

 \subsection{Isospin Effects at the Quark End of the String}
\label{qend}

Below we elaborate further on the subject of the influence of the quark flavor (u or d) on final state proton and neutron spectra reported in Sec.~\ref{pp}. This is necessary in view of the possibly non-trivial character of these effects as will be demonstrated below, and also in order to estimate the uncertainty related to this component of our model which remains essentially unconstrained by experimental data.

	Eq.~(\ref{eq1}) has the great advantage that it can be directly applied to experimental $pp$ and $pC$ data points. However, in the framework of the DPM and GEM models used in the present paper, this equation is {\em exact} for net baryon (summed net proton and net neutron) spectra, but only {\em approximate} for net protons and net neutrons if the latter are taken separately. The reason for this is that as a consequence of isospin effects at the quark end of the string, Fig.~\ref{isoflip}, already in the projectile hemisphere the $pp$ collision is not fully equivalent to the collision of the incoming proton with a single carbon ($^{12}_{~6}C$) nucleon, which has an equal probability to be a proton or a neutron; only the latter collision would be the correct representative for the single collision contribution to the minimum bias $pC$ sample. The corresponding correction to Eq.~(\ref{eq1}) reads
\begin{equation}
\Delta\left(\frac{dn}{dy}\right) =
~ \frac{P(1)}{1-P(1)} ~\biggl( \frac{dn}{dy}(pp\rightarrow(p-\bar{p})X) -
 \frac{dn}{dy}(pN\rightarrow(p-\bar{p})X )\biggr)
\label{eq2}
\end{equation}
for net protons and similarly for net neutrons. Here $pN$ labels the single proton-nucleon collision addressed above. Evidently, the correction 
$\Delta(\frac{dn}{dy})$ 
cannot be obtained from experimental $pp$ data alone. Therefore we decided to compute it using the DPM model and include it, with opposite sign, into the simulated distributions 
presented in Sec.~\ref{res} below.

\begin{figure}[t]
\begin{center}
\hspace*{-1cm}
\includegraphics[width=7cm]{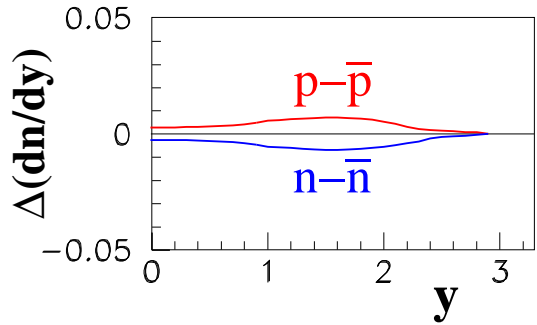}\\
\end{center}
\vspace{0.2cm}
  \caption{Correction $\Delta(\frac{dn}{dy})$ to $p-\overline{p}$ and $n-\overline{n}$ spectra
resulting from the difference in isospin effects between $pp$ and
$pN$ collisions as described in the text.}
 \label{isocor}
\end{figure}

	The computed correction 
$\Delta(\frac{dn}{dy})$ 
is quantified in Fig.~\ref{isocor} as a function of net proton (or net neutron) rapidity. The corresponding ($p-\overline{p}$) and ($n-\overline{n}$) distributions in $pN$ collisions, necessary for its evaluation, were simulated in a way identical to what was done for $pp$ reactions in Sec.~\ref{pp}. The only exception was the application of two new, different isospin flip functions, describing the isospin effects induced uniquely by the projectile proton quark and diquark, respectively. This was made because the ``nucleon'' target defined above was obviously isospin-symmetric, and brought no iso\-spin effects. The two functions are shown in Fig.~\ref{isoflip}, respectively as 
red and blue 
curve. As apparent in Fig.~\ref{isocor}, the correction 
$\Delta(\frac{dn}{dy})$ 
is opposite for net protons and net neutrons, which is an evident consequence of isospin symmetry. This correction is small, but it will nevertheless be taken into account in 
the considerations made in 
Sec.~\ref{res}.

Two remarks remain to be made at this point. Firstly, we 
simply stress the fact that our calculated value 
$\Delta(\frac{dn}{dy})$ 
is not zero at positive c.m.s. rapidity. This implies that 
once it is assumed - quite naturally in our view - that the flavor of the quark end of the string plays its role in the composition of particles resulting from its fragmentation,
the DPM will predict the presence of {\em isospin effects induced by the nucleon from the opposite hemisphere} (even if the latter effects are small).
%
%
Secondly, we underline that Fig.~\ref{isocor} also quantifies what we consider as upper limit of uncertainty induced in our model predictions by this specific isospin effect, and which results from the fact that the quark end of our effective isospin flip function (Fig.~\ref{isoflip}) is, for the time being, unconstrained by the data. This uncertainty remains moderate, not exceeding 0.01 at $y\sim 0.5$.

 \subsection{The Alternative Mechanism for Baryon Stopping}
\label{alt}

\begin{figure}[t]
\begin{center}
\includegraphics[width=3.6cm]{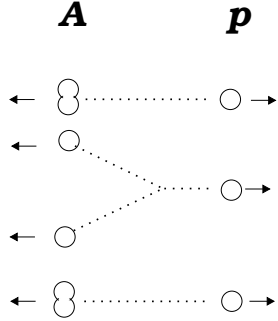}\\
\end{center}
  \caption{Basic diquark-disintegrating diagram which we introduce to the DPM as described in the text, shown for $pA$ collisions with $n=2$. 
The ``Y'' object, composed of three quarks with no diquark, is apparent in the center of the plot. This diagram was proposed by us in Ref.~\cite{g}.}
 \label{yobj}
\end{figure}

As our study made in Sec.~\ref{bstop} shows that the basic DPM mechanism for baryon stopping, namely the preservation and softening of the projectile diquark and the corresponding $D-q$ chain, cannot account for more than some two thirds of multiple collision events in $pC$ reactions, the question emerges what other mechanism is responsible for the remaining third of these events. Here we consider the most evident candidate which is the disintegration of the projectile diquark into separate valence quarks. The presence of such a process in multiple proton-nucleon collisions can be readily understood on the basis of simple color-based arguments as we demonstrated in Ref.~\cite{g}. For the present analysis we adopted the most evident diagram also mentioned therein, and presented in Fig.~\ref{yobj} for the case of $n=2$. Here the exchange of two gluons in the collision of the proton with two target nucleons results in the formation of three color singlets, each containing a single valence quark from the projectile. Two of these singlets are standard $q-D$ chains, evidently far more efficient in bringing the net baryon number in the direction of low or even negative rapidity than the standard softening of the $D-q$ chain. Evidently the fragmentation of these two chains
will be governed by the original fragmentation and isospin flip functions from Figs.~\ref{fragm} and~\ref{isoflip}, and thus it brings no new free elements to the model.
The third is a singlet of a new type, connecting one projectile and two target valence quarks. For the present study we take the somewhat risky assumption that the fragmentation of this ``Y object'' can be very different from a standard $q-D$ chain, and can be taken in fact as a free addition to the model. The consequences of taking this assumption will be discussed below.

\begin{figure*}[t]
\begin{center}
\hspace*{-1.0cm}
\includegraphics[height=8cm]{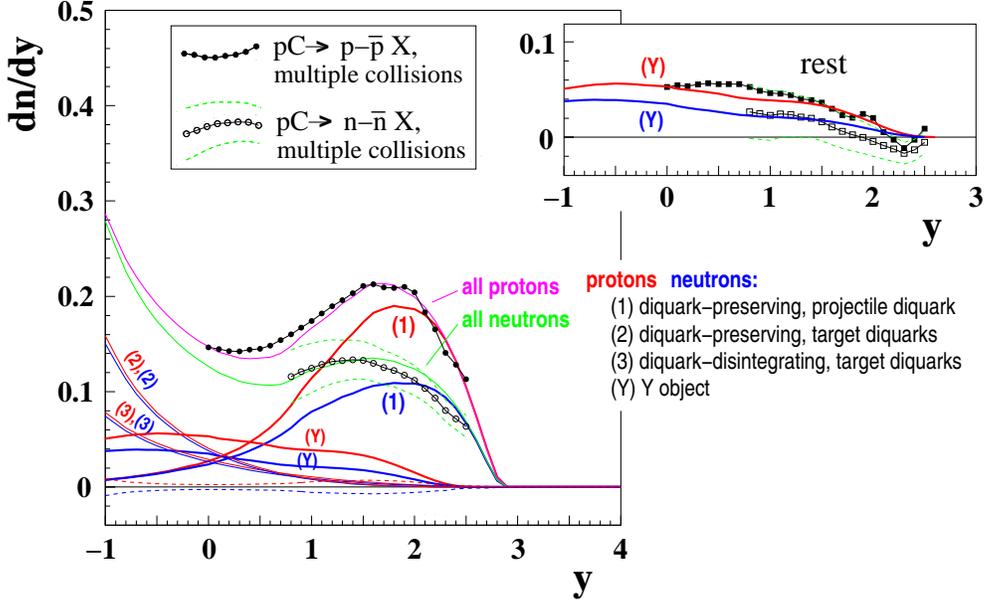}\\
\end{center}
 \caption{(Main figure:) distributions of net protons and net neutrons in $pC$ reactions in which the projectile proton undergoes more than one collision with carbon target nucleons, obtained from NA49 data~\cite{x,y} and compared to our DPM simulation 
described in the text. The different diquark-preserving and diquark-disintegrating 
contributions are indicated. The isospin correction from Sec.~\ref{qend} is included in the calculation of the total distribution and shown separately as dotted red and blue curves.
(Insert:) the ``rest'' distribution described in the text, compared to proton and neutron spectra from the ``Y object''. 
In both figures the green dashed curves reflect the systematic error of NA49 neutron data.}
 \label{pcmul}
\end{figure*}

The generalization of the diagram from Fig.~\ref{yobj} to $n=3, 4, ...$ proceeds in a way similar to that in the standard mechanism by Capella and Tran Thanh Van, by addition of $(n-2)$ quark-antiquark pairs from the sea. The distribution of valence and sea quarks in the projectile proton is taken from the formula we published in Ref.~\cite{g}:
\begin{multline}
\rho_m(x_{q_1},x_{q_2},x_{q_3},x_{1},...,x_{2m})
= \\
C_m
(x_{q_1}+x_{q_2})^{1/2}
x_{q_3}^{-1/2} 
\prod_{i=1}^{2m}(x_{i}^2+4\mu^2/s)^{-1/2} 
\cdot
\delta\Bigg( 1 - x_{q_1} - x_{q_2} - x_{q_3} - \sum_{i=1}^{2m}x_i\Bigg) ~~~~~~
\label{eq5}
\end{multline}
where $C_m$ is a normalization factor, 
${x_{q_1}}$, 
${x_{q_2}}$, 
${x_{q_3}}$, 
${x_{1}}$, 
               ..., 
${x_{2m}}$ 
are the momentum fractions carried by the three valence quarks and the $2m$ sea quarks and antiquarks (in the case considered here $2m=2n-4$), $\mu$ is the sea quark's transverse mass, and $s$ is the square of collision c.m.s.~energy. As we noted therein, 
Eq.~(\ref{eq5}) is a straight-forward generalization of the 
quark/diquark 
distributions from Eq.~(2.1a) in Ref.~\cite{MJ1985}. No modification of the sea quark's transverse mass $\mu$ from its original value of 0.3~GeV/$c^2$ was brought.
%
%

Following the considerations made above, the baryon stopping process in a collision of the projectile proton with more than one nucleon appears as a mixture of two processes, one preserving and one disintegrating the diquark. Our simulation includes both processes assuming a fixed probability for each of them independently on the number of proton-nucleon collisions $n$ ($n\geq 2$). We note that the latter assumption is non-trivial and in fact, some kind of $n$-dependence seems to us natural to consider here. However, at least for $pC$ collisions analyzed in this study, the steep, rapidly decreasing distribution of Glauber probabilities $P(n)$, see Ref.~\cite{pcdiscus}, inclines us to neglect this problem as far as the present paper is concerned\footnote{A better understanding of this issue can be brought by the analysis of proton-nucleus data with controlled centrality.}.

\subsection{Results}
\label{res}

Fig.~\ref{pcmul} shows the result of our best tuning of the DPM simulation to rapidity distributions of net protons and net neutrons in $pC$ collisions where the projectile proton collides with more than one target nucleon, 
obtained from experimental data through Eq.~(\ref{eq1}) and shown before in Fig.~\ref{pcpn}.
This tuned simulation corresponds to probabilities of diquark preservation and disintegration of 0.66 and 0.34, respectively ({\em i.e.}, 
the diquark-preserving component remains at its upper limit which we set 
up 
in Sec.~\ref{bstop}). The fragmentation and isospin flip functions of the ``Y object'' are taken as adjustable elements of the model and will be discussed below.

The overall description of the $p-\overline{p}$ and $n-\overline{n}$ distributions by the model appears satisfactory, even if the (tuned) simulation has a tendency to overestimate both proton and neutron yields at high and underestimate the proton yield at lower rapidity. The insert in Fig.~\ref{pcmul} shows the comparison between the simulated proton and neutron distributions obtained from the pure fragmentation of the ``Y object'' and the corresponding ``rest'' from the data points, obtained by subtracting all the simulated components of the model with the exception of the latter ``Y object'':
\begin{multline}
\frac{dn}{dy}\left(rest\right) ~=~
\frac{dn}{dy}\left(pC_\mathrm{multiple~collisions}\right)  
~-~ 
\frac{dn}{dy}\left(model\right)_\mathrm{~D~preservation,~all~chains} \\
~-~ \frac{dn}{dy}\left(model\right)_\mathrm{~D~disintegration,~q-D~chains} ~~~~~~~~.~~~
\label{eq3}
\end{multline}
%
%
From the above  we see that the simulated (and adjusted) fragmentation of the ``Y object'' matches the main characteristics of 
this
``rest'', once account is taken of the uncertainties induced by the errors of the experimental data. Still, we note that some tension remains present in this comparison. 

\begin{figure}[t]
\begin{center}
\hspace*{-1cm}
\includegraphics[width=6.5cm]{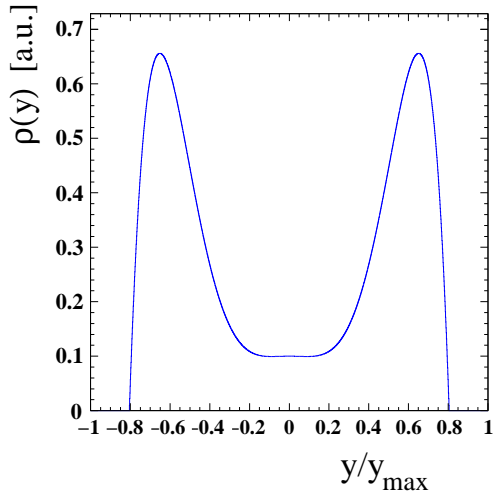}\\
\vspace*{-0.1cm}
\end{center}
  \caption{Fragmentation function of the ``Y object'' into baryons, drawn {\em versus} the rapidity ratio $y/y_\mathrm{max}$, where $y$ is the rapidity of the secondary baryon $B$ and $y_\mathrm{max}$ is the kinematical limit for $B$
(both $y$ and $y_\mathrm{max}$ are taken in the c.m.s.~of the ``Y~object'').}
 \label{9X}
\end{figure}

The overall success of our DPM simulation in the description of the very precise NA49 data (in $pp$ and ``multiple'' proton-nucleon collisions) will be somewhat diminished after inspection of the adjusted free element of the model, which is the fragmentation of the three-quark ``Y object''. The corresponding fragmentation and isospin flip functions which result in the best description of the data (shown in Fig.~\ref{pcmul}) are shown in Figs.~\ref{9X} and~\ref{10X}, respectively. The obtained fragmentation function of the ``Y object'', displays a characteristic symmetric, double-peaked shape, completely different from that of a standard $D-q$ chain presented in Fig.~\ref{fragm}. Of equal importance, the isospin flip function exhibits a very strong isospin effect at the proton valence quark end of the string, far larger than these we deduced for the $D-q$ chain in Fig.~\ref{isoflip}. Such a strong isospin effect (stronger than what we obtained for the diquark in the $D-q$ chain on the basis of $pp$ data, see Fig.~\ref{isoflip}) is in our view at the limit of conceivability as the resulting proton (or neutron) probability in the vicinity of the proton valence quark approaches the probability of the latter to be a u (or d).

 \subsection{Summary and Discussion of DPM Analysis}
\label{sumdpm}

The analysis presented in this section was in fact a direct application of the Dual Parton Model, in its formulation dedicated to baryon studies from Ref.~\cite{MJ1985}, to modern and far more complete experimental data~\cite{x,y}. No significant modification was brought to the methodology, and the only modified elements were the (effective) chain fragmentation and isospin flip functions which were adjusted to experiment as postulated in Ref.~\cite{MJ1985}. As such, this analysis was a completion of the study which we started in our earlier paper~\cite{e}.

\begin{figure}[t]
\begin{center}
\vspace*{0.2cm}
\hspace*{-1cm}
\includegraphics[width=6.6cm]{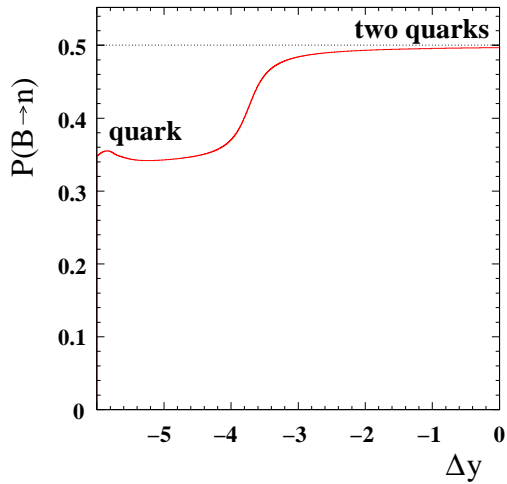}\\
\end{center}
  \caption{Isospin flip function which gives the probability $P(B\rightarrow n)$
of the non-strange baryon to become
a neutron in the fragmentation of the ``Y object''. 
Note: for clarity this function is plotted in the same way as in Fig.~\ref{isoflip}, with the two quarks taken instead of the diquark. As the latter two quarks belong to nucleons from the isospin-symmetric carbon nucleus, isospin effects emerge only from the one-quark side.}
 \label{10X}
\end{figure}

This approach appeared capable of providing a good description of existing $pp$ proton and neutron data in the ``non-diffractive'' region. Consequently, it pointed at the importance of a proper treatment of isospin effects induced by the diquark/quark flavors at both ends of the string. 

It also clearly demonstrated the impossibility to describe baryon spectra in ``multiple'' proton-nucleon collisions solely on the basis of the ``standard'' mechanism of baryon stopping postulated by Capella and Tran Thanh Van, which in our view forces the interpretation of very frequent disintegration of the projectile diquark. The upper limit for the diquark-preserving ``standard'' mechanism appeared to be of the order of two thirds. 

Finally, the attempt of obtaining a {\em complete} description of the above baryon spectra by introducing a new, diquark-disintegrating contribution was essentially successful (Fig.~\ref{pcmul}), but at the price of accepting the new element of this process, the ``Y~object'' made up of three quarks with no diquark, to have rather exotic fragmentation characteristics, namely very strong isospin effects at the proton quark end, Fig.~\ref{10X}, and a symmetric, double-peaked fragmentation function, Fig.~\ref{9X}.

The above summarizes the status of understanding of the baryon stopping processes which we reached with the DPM. We still note that the double-peaked structure obtained in Fig.~\ref{9X} was at the source of 
our \mbox{intuition}
that the basic diquark-disintegrating contribution is in fact made not by {\em one}, but {\em two} partially symmetric diagrams which we in fact introduced as (e) and (f) in Ref.~\cite{g}. Consequently the question emerged whether a more general model, free from the necessity of accepting exotic or unnatural chain fragmentation schemes, could be developed in order to provide a good description of the experimental $pp$ and $pC$ data~\cite{x,y}. The present status of our work on such a model will be described in the next Section.

 \section{The Gluon Exchange Model Analysis}
\label{gemana}

In this section we discuss the analysis made with our new Gluon Exchange Model (GEM) which we recently introduced in Ref.~\cite{g}. As this latter model was already presented in some detail therein, we will only shortly summarize the already presented findings.
We will mostly concentrate on new results not included in Ref.~\cite{g}.

\begin{figure*}[t]
\begin{center}
\includegraphics[width=11.4cm]{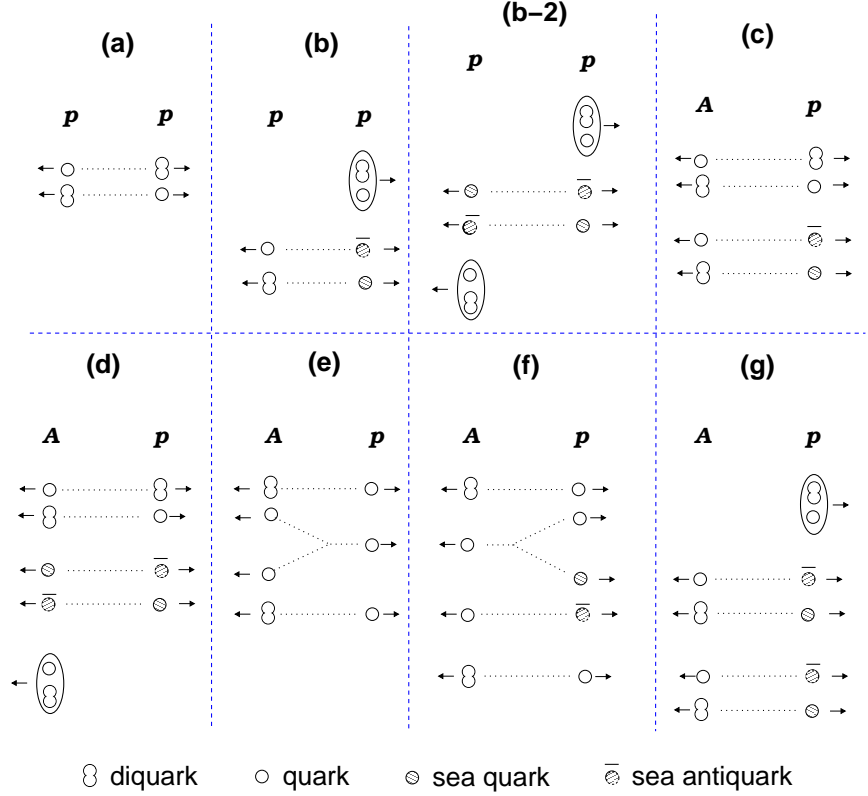}\\
\end{center}
\vspace*{0.2cm}
  \caption{List of basic diagrams included in GEM.
All the diagrams corresponding to $pA$ collisions (c-g) are drawn for the case of the proton projectile colliding with $n=2$ nucleons.
A similar 
way of plotting all the diagrams but (b-2) was 
earlier
introduced by us in Ref.~\cite{g}.}
 \label{diag}
\end{figure*}

\subsection{GEM}
\label{gem}

GEM is a model designed for a {\em full}, homogeneous description of hadron-hadron and hadron-nucleus reactions based on color octet (gluon) exchange. The model was built upon the following postulates:

\begin{enumerate}
\item
the number of hadron-nucleon collisions in the hadron-nucleus reaction is given by the number of exchanged color octets;
\item
a homogeneous description of the entire baryon spectrum is possible with a properly complete Fock space of states for the participating protons and nucleons;
\item
the fragmentation of {\em any} chain with non-zero baryon number (that is, made either of a diquark and quark, or of three quarks) is similar to that of a standard $D-q$ chain; in other words exotic fragmentation functions of the type 
shown in Fig.~\ref{9X} are not permitted by the model.
\end{enumerate}
%
We underline that the version of GEM presented in Ref.~\cite{g} and in the present paper should still be considered as a 
``rough and ready'' or ``effective''
approach to this model. What is meant here is that 
this version 
is ``minimalistic in implementation'' - it inherits as much of the original formalism from the Dual Parton Model described in Refs.~\cite{CapellaTranh,MJ1985} as it is allowed by experimental data. Specifically,
the fragmentation- and isospin-related elements of the model still fully follow the methodology 
proposed in Ref.~\cite{MJ1985} and applied for the DPM in Sec.~\ref{dpm}. The comparison of the corresponding fragmentation and isospin flip functions will be given 
below. It is evident for us that other, more rigorous approaches to GEM
are possible although this subject is beyond the scope of this paper.

The general parton momentum distribution in GEM is defined by Eq.~(\ref{eq5}) given  in Sec.~\ref{alt} above.
In Fig.~\ref{diag}, we list the basic diagrams considered by GEM for proton-proton and proton-nucleus collisions. All the diagrams but (b-2) have  already been 
explicitly
introduced in Ref.~\cite{g} and are reminded here only for clarity\footnote{The configuration (b-2) was discussed in the text of Ref.~\cite{g}.}. Attention should be drawn upon the extension of the Fock space of available states which materializes in new diagrams where color octet exchange couples to one, or more, quark-antiquark pairs from the sea, leaving intact the valence color singlet $Dq$ (diagrams b, b-2, d, and g), and to the presence of 
the
diquark-disintegrating diagrams (e) and~(f)
which are partially symmetric to each other.
 It is evident that for some of the diagrams shown, the natural extension to a higher number of collisions by addition of further $(q_s,\overline{q}_s)$ pairs, as well as appropriate combinatorics are taken into account in our calculations.

\begin{figure}
\vspace{0.2cm}
\begin{center}
\hspace*{-1cm}
\includegraphics[width=6.5cm]{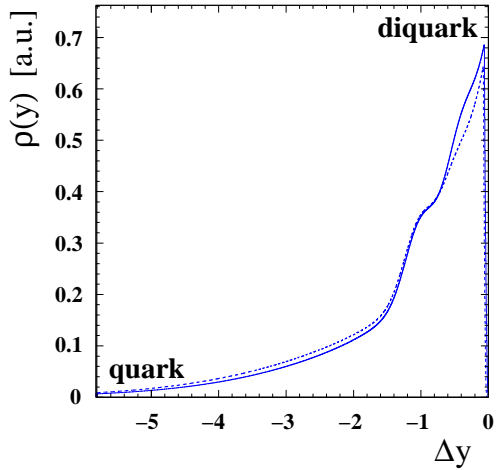}\\
\end{center}
  \caption{Comparison of fragmentation functions of the $D-q$ (diquark-quark) chain into baryons, used in our DPM (solid) and GEM (dashed) simulation. Both functions are drawn {\em versus} the 
difference 
$\Delta y=y-y_\mathrm{max}$, where $y$ is the rapidity of the secondary baryon $B$ and $y_\mathrm{max}$ is the kinematical limit for $B$ in the direction of $D$.}
 \label{12x}
\end{figure}

\begin{figure}
\vspace{0.2cm}
\begin{center}
\hspace*{-2.4cm}
\includegraphics[width=8.1cm]{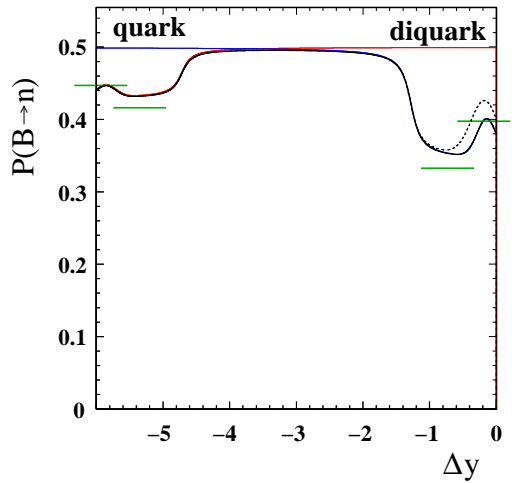}\\
\end{center}
  \caption{Comparison of isospin flip function for the $D-q$ chain, used in our DPM (solid) and GEM (dashed) simulations.}
 \label{13x}
\end{figure}

The main findings achieved with GEM already reported in Ref.~\cite{g} can be summarized as follows:

\begin{enumerate}
\item[$\bullet$] 
GEM provides a homogeneous description of the {\em entire} proton and neutron spectra in $pp$ collisions. In this sense this description is far better than that obtained with DPM in Sec.~\ref{pp} as it includes the proton ``diffractive peak'' at high rapidity. This is achieved by the addition of the diagram (b) from Fig.~\ref{diag} which brings the emission of such fast $Dq$ (uud) singlets as a result of color octet exchange with the proton in its {\em next} Fock state including one  $(q_s,\overline{q}_s)$ pair. This challenges the long-standing opinion that ``diffractive'' protons belong to another regime~\cite{9x} and 
that their emission is
not caused by color exchange.
\item[$\bullet$] 
The shape of the experimental ``diffractive peak'' imposes a severe constraint on the probability of diagram (b) as well as on the mass of the sea quark $\mu$, with $\mu=60$~MeV/$c^2$ giving the best description of experimental data.
\item[$\bullet$] 
Similarly as this was done for the DPM in Sec.~\ref{pp}, the $D-q$ chain fragmentation and isospin flip functions are tuned to the experimental $pp$ data. The resulting functions are shown in Figs.~\ref{12x} and~\ref{13x}, respectively. As apparent in the figures, the modification with respect to the functions obtained from the DPM analysis is in fact very moderate.
\item[$\bullet$] 
Finally, the application of GEM to proton-carbon reactions in which the projectile proton collides with more than one target nucleon (see Fig.~\ref{pcpn} and Eq.~\ref{eq1}) shows that the diquark-preserving diagrams alone
are not sufficient to provide any description of the corresponding proton and neutron spectra;
in fact they
clearly fail to describe the experimental data\footnote{The analysis made in Ref.~\cite{g} included the diquark-preserving diagrams (c) and (d), as the diagram (g) is strongly suppressed.}. 
As such, this study leads to the same conclusion as for the DPM, that the diquark must be frequently disintegrated in such processes. The corresponding upper limit for the diquark-preserving mechanism is about a half, which is of the same order but more restrictive than what we obtained for the DPM in Sec.~\ref{bstop}. 
\end{enumerate}

In the following section we will summarize our first results on the inclusion of basic diquark-disintegrating diagrams into GEM.

 \subsection{Transport of Baryon Number  in Proton-Nucleus Reactions in GEM}
\label{bsgem}

\begin{figure}[t]
\begin{center}
\hspace*{-1.4cm}
 \includegraphics[height=8cm]{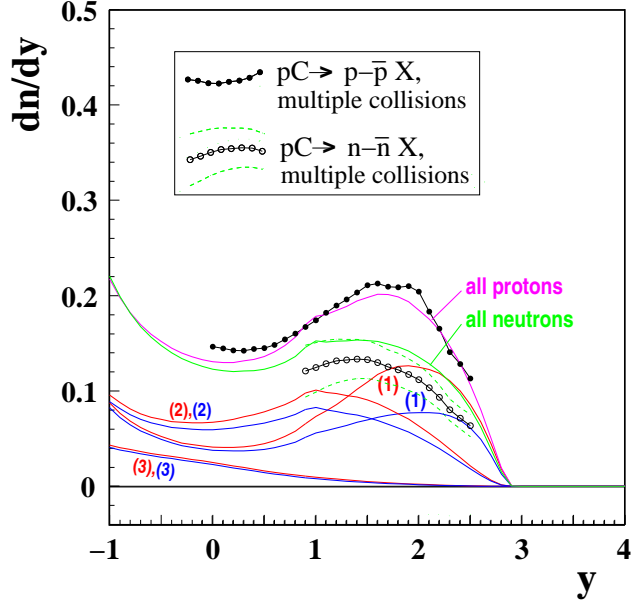}\\
\end{center}
 \caption{Distributions of net protons and net neutrons in $pC$ reactions in which the projectile proton undergoes more than one collision with carbon target nucleons, obtained from NA49 data~\cite{x,y} and compared to our GEM simulation 
described in the text. The different contributions to the proton distribution are marked in red and to the neutron distribution are marked in blue: (1) diagrams c+d, (2)~diagram~f, (3) diagram e. The isospin correction from Sec.~\ref{qend} is included in the calculation of the total simulated proton and neutron distribution.
                The green dashed curves reflect the systematic error of NA49 neutron data.}
 \label{mulgem}
\end{figure}

Fig.~\ref{mulgem} gives a full summary of our study of the multiple proton-nucleon process in $pC$ reactions made with our present 
version of GEM. The study is made on the following premises:
\begin{enumerate}
\item[(i)]
Four diagrams are included in the study. These are the diquark-preserving diagrams from Ref.~\cite{g} (c, d), taken together as it was made therein, and the two new diquark-disintegrating diagrams (e) and (f). The strongly suppressed diagram (g) is neglected as in Ref.~\cite{g}.
\item[(ii)]
The fragmentation and isospin flip function of all the created $D-q$ and $q-D$ chains are taken from $pp$ collisions as shown in Figs.~\ref{12x} and~\ref{13x}, respectively. What is more, {\em also for the diquark-disintegrating diagrams} the fragmentation and isospin flip functions of color singlets made of three quarks with no diquark (``Y objects'' in Fig.~\ref{diag} e, f) are taken following the same prescription. This implies that we assume 
that the two quarks from the target (in diagram e) and projectile (in diagram f) form a kind of ``effective diquark'', which is justifiable on the basis of color-based arguments.
\item[(iii)]
Finally, similarly to what was done in Sec.~\ref{res}, the correction for isospin effects at the quark end of the string (Eq.~\ref{eq3}) is also applied in our simulation. We note that we apply the same correction as obtained from our DPM studies in Sec.~\ref{res}. The corresponding relative uncertainties are of the order of changes in fragmentation and isospin flip functions between the two analyses, and therefore can be neglected account taken of the small size of this correction. We also note that this correction actually {\em worsens} the agreement between data and model, 
and that the corresponding uncertainty induced in our study was already discussed in Sec.~\ref{qend}.
\end{enumerate}
%

The respective probabilities which provide the best simultaneous description of proton and neutron spectra are 0.46 for the summed diquark-preserving diagrams (c+d), 0.12 for diagram (e) and 0.42 for diagram (f). We conclude that this description is satisfactory account taken of the 
uncertainties attributable to experimental data~\cite{x,y} and, most of all, of the restrictiveness of the model, see item (ii) above. It is clear that formulated as above, GEM has no difficulty in providing the correct amount of transport of the {\em total} projectile baryon number that would correspond to net protons and neutrons put together as it was done in Ref.~\cite{e}. 

What is less clear is to what extent a successful description can be provided for $p-\overline{p}$ and 
$n-\overline{n}$ spectra taken {\em separately}, as shown in Fig.~\ref{mulgem}. 
Here the present version of the model overestimates neutron and underestimates proton emission. The discrepancy is {\em increased} by the correction for isospin effects at the quark end of the string which, as it was said in Sec.~\ref{qend}, introduces some uncertainty in our study. This part of the model may be subject to improvement in the framework of future studies.

To sum up, 
our present
approach to the GEM model allows for an exact description of net proton and neutron spectra in $pp$ collisions, and a reasonable description of $pC$ reactions including these in which the projectile proton collides with more than one nucleon. This is achieved with a restrictive model which does not involve any new fragmentation schemes for the new color singlet configurations (three-quark objects) created in multiple collisions, thus in this respect this description is better than the one obtained in Sec.~\ref{res} for the Dual Parton Model with an {\em ad hoc} Y-object fragmentation function. 
However, while the mere transport of baryon number in multiple collision processes does not seem to be a major problem for the present version of GEM, the isospin effects which govern the difference between proton and neutron emission call for attention and further development of the model. As these effects appear strong for forward baryon emission at CERN SPS energies and at the same time, they may provide evidently valuable information on the fate of quarks participating in the collision, further theoretical and experimental effort in this direction is clearly indicated.

\section{Summary and Discussion}
\label{sum}

This paper served as a summary of the present status of our (ongoing) phenomenological studies of the transport of baryon number in proton-proton and proton-nucleus collisions at CERN SPS energies. This energy regime proved optimal by virtue of (1) low baryon-antibaryon pair production while at the same time the collision remains in the high energy regime (2) strong isospin effects on final state baryon spectra and (3) the availability of modern experimental data~\cite{x,y} on both protons and neutrons in nearly the entire projectile hemisphere of the reaction. Numerous elements of this study were published before in Refs.~\cite{e} and~\cite{g}.
This paper is third in a row and is meant to complete the information presented therein. 

This study was made using two parton models, the classic Dual Parton Model and our new Gluon Exchange Model with significantly extended Fock space. Below we will attempt a synthetic discussion of the main items which emerge from our analysis.

\begin{enumerate}
\item[$\bullet$]
The availability of modern experimental data allows for a significant progress in understanding of the process of transport of baryon number. This is mostly due to the eliminated necessity of using protons alone as a proxy for the total baryon number, nearly full coverage with well controlled systematic errors which allows the use of baryon number conservation as an additional tool for verification of the model, and the possibility of extraction of multiple proton-nucleon processes in proton-nu\-cleus collisions. As a result, phenomenological conclusions more precise than in earlier studies~\cite{MJ1985,MJZalewski} can be obtained.
\item[$\bullet$]
A natural extension of the Fock space of states available for participating protons allows for an accurate description of the entire proton and neutron distribution in $pp$ collisions. Consequently, GEM reaches a more complete description of the latter than DPM, and explains the emission of so-called diffractive protons at high $x_F$ as a specific case of color octet exchange which challenges earlier ideas on this subject~\cite{9x}.
\item[$\bullet$]
Independently on the model applied and independently on studied observable (summed non-strange baryons~\cite{e}, protons, neutrons), the standard diquark-preserving mechanism of baryon stopping in $pA$ reactions proposed by Capella and Tran Thanh Van fails to reproduce the spectra of baryons in $pC$ collisions in which the proton projectile interacts with more than one nucleon. This way the NA49 data~\cite{x,y} force us to conclude that the diquark in the projectile proton, if it exists, must be very frequently disintegrated already in collisions with light $C$ nuclei. 
\item[$\bullet$]
The upper limit which we obtain for diquark-preserving processes in such collisions varies from a half to two thirds depending on the model applied. It is clearly conceivable that this upper limit will go further down with increasing size of the target nucleus, for instance for $pPb$ reactions. 
This calls for scrutiny of other models involving the diquark in soft hadronic or nuclear processes~\cite{bialasbzdak1,bialasbzdak2,bialasbzdak3}.
\item[$\bullet$]
After inclusion of diquark-disintegrating scenarios, both models achieve an essentially reasonable description of baryon stopping in multiple proton-nucleon collision processes in $pC$ reactions, but with important provisos. The satisfactory description we obtained with DPM went at the price of postulating quite an exotic fragmentation scenario, including among others very strong isospin effects, for the new three-quark color singlet configuration appearing in such collisions. GEM, on the other hand, permits such a description without applying such exotic scenarios but seems to underestimate the strong isospin effects emerging from experimental data. As these effects are informative on the fate of quarks in the collision this subject calls for further scrutiny.
\item[$\bullet$]
Finally, our studies bring interesting implications in view of possible future measurements. Apart from proton-nu\-cleus and meson-nucleus collisions involving heavier ato\-mic nuclei, a particularly interesting possibility are {anti\-proton-nucleus} reactions\footnote{See Ref.~\cite{Bailey} for comparison.} where, as a consequence of disintegration of the projectile anti-diquark, multi-step annihilation diagrams for projectile anti-quarks on multiple target nucleons should show up as a function of collision centrality. A first proposal for such studies has already been communicated to the NA61/SHINE Collaboration~\cite{s}. Other research programs, including in particular AFTER@LHC, are also under consideration.
\end{enumerate}

\vspace*{0.3cm}
\noindent
{\Large\bf Acknowledgements}\\
\vspace*{-0.1cm}

\noindent
This work was supported by the National Science Centre, Poland (grant no.
2014/14/E/ST2/00018).


\end{document}